\documentclass[a4paper,10pt]{amsart}
\usepackage{graphicx}
\usepackage{amssymb,amsmath,amstext,amscd}

\theoremstyle{plain}
\newtheorem{thm}{Theorem}[section]

\newtheorem*{main theorem}{Theorem}

\theoremstyle{definition}

\begin{document}

\title{A simple proof of Zeeman's theorem}
\author{Do-Hyung Kim}
\address{Department of Mathematics, College of Natural Science, Dankook University,
San 29, Anseo-dong, Dongnam-gu, Cheonan-si, Chungnam, 330-714,
Republic of Korea} \email{mathph@dankook.ac.kr}


\begin{abstract}
It is discussed that Zeeman's theorem can be directly obtained
from Liouville's theorem if we assume sufficient
differentiability.
\end{abstract}

\maketitle

\section{Introduction} \label{section:1}

 In 1964, Zeeman has shown that causal isomorphism on Minkowski
 space $\mathbb{R}^n_1$ is generated by dilatation, translation
 and orthochronous matrix.(See Ref. \cite{Zeeman}) In his theorem, he assumed neither
 differentiability nor continuity of causal isomorphism and so the theorem tells
 us that if a bijection preserves causal relation, then the map
 becomes smooth diffeomorphism. However, as he remarked, his
 theorem does not hold if $n=2$.

 In this paper, we show that if a smooth diffeomorphism preserves
 causal relation, Zeeman's result can be obtained by simple
 argument of applying known results. This argument explains
 explicitly the reason why Zeeman's theorem does not hold in
 two-dimensional Minkowski space. The same idea as in this paper
 was suggested by authors in Ref.\cite{Minguzzi}.

\section{A new proof} \label{section:2}

We define a smooth map $F : \mathbb{R}^n_1 \rightarrow
\mathbb{R}^n_1$ to be a causal isomorphism if it is bijective and
satisfies the condition that $x \leq y$ if and only if $F(x) \leq
F(y)$.

Then, by Hawking's theorem(Ref. \cite{Hawking}), or by Proposition
3.13 in Ref.\cite{Ming} $F$ is a smooth conformal diffeomorphism
and then we can apply the following theorem of Liouville.

\begin{thm}
Let $U$ be an open subset of semi-Euclidean space
$\mathbb{R}^n_\nu$ with $n \geq 3$ and $F : U \rightarrow
\mathbb{R}^n_\nu$ be a conformal diffeomorphism. Then, $F$ is a
composite of isometry, dilatation and inversion.
\end{thm}

 Elementary proof of this theorem with the assumption that $F$ is $C^4$ can be found in
 Ref.\cite{Blair} and Theorem 15.2 in Ref.\cite{DFN}.

We remark that Liouville's theorem holds when $n \geq 3$ and does
not hold for $n=2$. In fact, any holomorphic map defined on an
open subset of $\mathbb{R}^2$ with nonzero derivative, is
conformal.

To apply Liouville's Theorem, we note that inversion has a
singularity and so if $F : \mathbb{R}^n_1 \rightarrow
\mathbb{R}^n_1$ is a conformal diffeomorphism defined on the whole
of $\mathbb{R}^n_1$, we must discard inversions. Therefore, if $F
: \mathbb{R}^n_1 \rightarrow \mathbb{R}^n_1$ is a smooth causal
isomorphism, then $F$ is a conformal diffeomorphism and by
Liouville's theorem, $F$ is a composite of isometry, dilatation
and inversion. However, since $F$ is defined on the whole of
$\mathbb{R}^2_1$, we must discard inversion and so $F$ is a
composite of dilatation and isometry. In conclusion, we have the
following.

\begin{thm}
Let $F : \mathbb{R}^n_1 \rightarrow \mathbb{R}^n_1$ be a $C^1$
causal isomorphism with $n \geq 3$. Then we have $F(\mathrm{x}) =
a A \mathrm{x}+b$ where $a>0$ and $A$ is an orthochronous matrix.
\end{thm}

As we can see in Proposition 3.13 in Ref.\cite{Ming}, any smooth
causal isomorphism on $\mathbb{R}^n_1$ is a conformal
diffeomorphism regardless of its dimension $n$. However, since
Liouville's theorem does not hold for $n=2$, we obtain Zeeman's
theorem only for $n \geq 3$.

\section{Acknowledgement}

The present research was conducted by the research fund of Dankook
University in 2013.

\end{document}